\documentclass[english,aps,preprint]{revtex4}
\usepackage[T1]{fontenc}
\usepackage[latin9]{inputenc}
\usepackage{array}
\usepackage{multirow}
\usepackage{amsmath}
\usepackage{graphicx}

\makeatletter

\providecommand{\tabularnewline}{\\}

\@ifundefined{textcolor}{}
{%
 \definecolor{BLACK}{gray}{0}
 \definecolor{WHITE}{gray}{1}
 \definecolor{RED}{rgb}{1,0,0}
 \definecolor{GREEN}{rgb}{0,1,0}
 \definecolor{BLUE}{rgb}{0,0,1}
 \definecolor{CYAN}{cmyk}{1,0,0,0}
 \definecolor{MAGENTA}{cmyk}{0,1,0,0}
 \definecolor{YELLOW}{cmyk}{0,0,1,0}
}

\makeatother

\usepackage{babel}
\begin{document}

\title{Single production of composite electrons at the future SPPC-based
lepton-hadron colliders }

\author{A. Caliskan}

\email{acaliskan@gumushane.edu.tr}

\affiliation{Gümü\c{s}hane University, Faculty of Engineering and Natural Sciences,
Department of Physics Engineering, 29100, Gümü\c{s}hane, Turkey}
\begin{abstract}
We consider the production of excited electrons with spin-1/2 at the
future SPPC-based electron-proton colliders with center-of-mass energies
of $8.4$, $11.6$, $26.6$ and $36.8$ TeV. These exotic particles
are predicted in the composite models. We calculate the production
cross-sections and concentrate on the photon decay channel of the
excited electrons with the process of $ep\rightarrow e^{\star}X\rightarrow e\gamma X$
. The pseudorapidity and transverse momentum distributions of the
electrons and photons in the final-state have been plotted in order
to determine the kinematical cuts best suited for discovery of the
excited electrons. By applying these cuts we compute $2\sigma$, $3\sigma$
and $5\sigma$ contour plots of the statistical significance of the
expected signal in the parameter space ($L$, $m^{\star}$), where
$L$ denotes the integrated luminosity of the collider and $m^{\star}$
is the mass of the composite electrons.
\end{abstract}
\maketitle

\section{introduction}

Elementary particles and electromagnetic, weak and strong interactions
among them today are well described by the Standard Model (SM) in
particle physics. Discovery of the Higgs boson with a mass of approximately
125 GeV by the ATLAS \cite{observation-ATLAS} and the CMS \cite{observation-CMS}
collaborations in 2012 confirmed the electroweak symmetry breaking
mechanism of the SM. This discovery has further increased the reliability
and accuracy of the SM. But, the SM still contains many puzzles and
will likely need some modifications in the future. Some of the main
problems to be solved in the SM are the neutrino masses, quark-lepton
symmetry, family replication, large number of free parameters, CP
violation, fermion's masses and mixing pattern. The most powerful
candidates that can solve these problems are theories beyond the SM
that have been studied over the years. A lot of theory like Technicolour
\cite{weinberg,suskind}, Grand Unified Models \cite{unity of all,lepton number}
and Supersymmetry \cite{extension of the} are proposed so far, but
the most important one is the compositeness \cite{souza} that better
explains the proliferation of elementary particles, introducing new
fundamental constituents called preons.

Lepton and quark compositeness were firstly proposed at the 1970s
\cite{unified model,subquark model,observable effects,a fundamental theory}
, and so far many preonic models claiming that quarks and leptons
have a sub-structure are proposed. The most important ones of them
are the Fritzsch-Mandelbaum model \cite{weak interactions,q quantum structure}
(the so-called haplon model), Harari-Shupe model \cite{a schematic model,shupe}
(the so-called rishon model), Terazawa WCH model \cite{unified model}
and Abboth-Farhi \cite{abbothh-Farhi} model. In the Fritzsch-Mandelbaum
model, for example, leptons and quarks are bound states of four fundamental
preons, two of which are fermionic and two are bosonic, called $\alpha$,$\beta$,$x$
and $y$. Possible new interactions between the fermions should occur
at the scale of binding energies that connect the preons togetter.
This energy scale is an important parameter for all composite models,
and called compositeness scale, $\varLambda$.

As a consequence of the compositeness we expect that the SM leptons
will have their excited states if they have a sub-structure. Therefore
the excited leptons ($e^{\star},\mu^{\star},\tau^{\star},\nu_{e}^{\star},\nu_{\mu}^{\star},\nu_{\tau}^{\star}$)
are studied by the composite models. It is estimated that the masses
of the excited leptons, which may have spin-1/2 or higher spin states,
need to be heavier compared to the SM leptons.

In the literature there are important recent searches that have studied
the production of excited leptons in various colliders and estimated
mass limits for their discovery. Composite majorana neutrinos \cite{hunting for}
and doubly charged excited leptons \cite{phenomenology,double charged}
have been studied in detail at the Large Hadron Collider (LHC) energies.
When we look at the future colliders, the studies of excited neutrino
production at the CLIC \cite{analysis of excited} and the LHeC \cite{search for}
colliders, and excited muon \cite{excited muon searches} and neutrino
\cite{excited neutrino search} productions at the FCC-based future
colliders have been performed. Although several theoretical and phenomological
studies have been carried out on the composite leptons, no evidence
has yet been obtained about the existence of them in the experimental
studies at the LEP \cite{search-LEP}, HERA \cite{search-HERA}, TEVATRON
\cite{search-DO}, ATLAS \cite{search for excited electrons} and
CMS \cite{search for excited leptons}.

The greatest benefit of the experimental studies is the exclusion
of energy regions where there is no signal for the existence of the
excited leptons, and consequently the determination of a lower mass
limit for the subsequent studies. The most up-to-date experimental
results for the excited electrons were obtained from the OPAL \cite{PDG}
and CMS \cite{search for excited leptons} experiments carried out
at the CERN. In the OPAL experiment, the pair production of the excited
electrons was investigated by electron-positron collisions ($e^{+}$$e^{-}$$\rightarrow$$e^{\star}$$e^{\star}$),
and the mass limit was determined as $m_{e^{\star}}$ > $103.2$ GeV.
The single production of the excited electrons ($pp\rightarrow ee^{\star}X$)
was searched for and the mass values up to $3900$ GeV were excluded.
These experimental mass limits for the excited leptons were obtained
from the gauge interactions assuming $f=f'=1$ and $\varLambda=m_{e^{\star}}$.

In addition to gauge interactions, another production mechanism of
the excited leptons are the four fermion contact interactions. It
is known that the contact interactions contribute to the production
of the excited leptons at a level comparable to their gauge interactions.
In fact, in a recent phenomenological study on proton-proton collisions
\cite{hunting for} it has been shown that the contact interactions
predominate for the production of the excited leptons at the LHC energies.
We have taken the gauge interactions into account for current work,
but it is not true that we omit the contribution of the contact interactions
altogetter. So we have planned the contact interaction version of
this work as the subject of a future study.

In this paper we have searched for the single production of the excited
electrons at the Super Proton-Proton Collider (SPPC)-based electron-proton
colliders including four different center-of-mass energy options.
After a general introduction in this section, the rest of the paper
is organised as follows: in Section 2 we discuss the SPPC-based electron-proton
collider options and their general accelerator parameters; in Section
3 we present the excited electron interaction Lagrangian, its cross-sections
and decay widths; in Section 4 we do signal and background analysis
to determine mass limits for discovery; finally Section 5 contains
the final discussions and conclusions.

\section{THE SPPC-BASED LEPTON-HADRON COLLIDERS}

With discovery of the Higgs particle at the CERN particle physics
has reached the Higgs era, but it is unclear whether this particle
is a fundamental scalar. The next step to be done is that the properties
of the Higgs particle should be examined in detail and the true internal
structure should be understood. For this purpose, the feasibility
studies have been initiated to establish a Higgs factory all over
the world.

The LHC collider is the world's most powerful particle collider built
to date. It is still in operation and will continue to operate until
the 2030s as part of its high-luminosity upgrade programme. In order
to examine primarly the properties of the Higgs particle, the design
of various collider projects planned to be established in the future
has been started. Some of them are categorized in the literature as
the LHC era colliders. Important international collider projects in
the LHC era are the ILC (International Linear Collider) with $\sqrt{s}$
=$0.5$($1$) TeV \cite{ILC-TDR}, low energy muon collider ($\mu^{-}\mu^{+}$)
\cite{muon-accelerator}, LHeC (Large Hadron Electron Collider) with
$\sqrt{s}$ =$1.3$ TeV \cite{LHeC-webpage} and CLIC (Compact Linear
Collider) with $\sqrt{s}$ =$3$ TeV (optimal) \cite{CLIC-CDR}. The
design works of these international projects have reached a certain
stage.

The most important collider project to be established in the Europe
for the post-LHC era is the international Future Circular Collider
(FCC) project \cite{FCC-webpage}. The design works of the FCC, which
will have a center-of-mass energy of 100 TeV, started at CERN in 2010-2013
and supported by the European Union within the Horizon 2020 Framework
for Research and Innovation. The main purpose of the FCC project to
be built in the 80-100 km new tunnel at the CERN is to install an
energy-frontier hadron-hadron collider (FCC-hh) with $\sqrt{s}$ =$100$
TeV. In the next phase of the project, it is also planned to establish
a lepton-lepton collider (FCC-ee or TLEP \cite{TLEP-webpage}) with
a center-of-mass energy of 90-400 GeV to the same tunnel. With the
lepton collider being activated the lepton-hadron collider option
(FCC-he) will be available as a third collider type. The FCC project
will allow us to search for the Higgs particle and new interactions
beyond the SM at the highest energies. The Conceptual Design Report
(CDR) of the project has been written as four volumes in 2018 \cite{FCC-CDR-1,FCC-CDR-2,FCC-CDR-3,FCC-CDR-4}.

Another important collider project to be established in the post-LHC
era is the CEPC-SPPC collider. The CEPC-SPPC which was started to
be designed by Chinese physicists in 2012 is a two-stage circular
collider project. Its Preliminary Conceptual Design Report (Pre-CDR)
was completed in 2015 \cite{SPPC-CDR} and the CDR of the project
was published in 2018 as two volumes \cite{SPC-CDR-1,SPC-CDR-2}.
In the first stage of the project according to this reports, an electron-positron
collider, called the Circular Electron Positron Collider (CEPC), with
a center-of-mass energy of 240 GeV will be install to investigate
primarly the Higgs physics in a 100 km tunnel. After the CEPC collider
completes its missions it will be upgraded to the second stage. In
the second stage an energy-frontier hadron-hadron collider, called
Super Proton Proton Collider (SPPC), with a center-of-mass energy
of more than 70 TeV will be installed in the same tunnel. Although
the center-of-mass energy of the SPPC collider is about 70 TeV according
to the Pre-CDR, this value was increased to 75 TeV in the CDR report.
But, the ultimate goal of the project is to reach higher center-of-mass
energy levels. Table 1 denotes the general parameters reported in
the Pre-CDR and CDR of the SPPC collider, including various design
options \cite{SPPC-parameter}. 

\begin{table}
\caption{The main parameter options of the proton beams for the SPPC collider.}

\begin{tabular}{|c|c|c|c|c|}
\hline 
Parameters  & Option-1 (Pre-CDR)  & Option-2  & Option-3 (CDR)  & Option-4 \tabularnewline
\hline 
\hline 
Beam energy (TeV)  & $35.6$  & $35$  & $37.5$ & $68$ \tabularnewline
\hline 
Circumference (km)  & $54.7$  & $54.7$  & $100$  & $100$ \tabularnewline
\hline 
Dipole field (T)  & $20$  & $19.69$  & $12$ & $20.03$ \tabularnewline
\hline 
Peak luminosity ($\times10^{35}cm^{-2}s^{-1}$)  & $1.1$  & $1.2$  & $1$  & $10.2$ \tabularnewline
\hline 
Particle per bunch ($10^{11})$  & $2$  & $2$  & $1.5$  & $2$ \tabularnewline
\hline 
Norm. transverse emittance ($\mu m$)  & $4.1$  & $3.72$  & $2.4$  & $3.05$ \tabularnewline
\hline 
Bunch number per beam  & $5835$  & $5835$  & $10080$  & $10667$ \tabularnewline
\hline 
Bunch length (mm)  & $75.5$  & $56.5$  & $75.5$  & $15.8$ \tabularnewline
\hline 
Bunch spacing (ns)  & $25$  & $25$  & $25$  & $25$ \tabularnewline
\hline 
\end{tabular}
\end{table}

A lepton-hadron collider option can be obtained if a linear electron
accelerator is installed tangentially to the SPPC proton complex.
It is important here that the electron accelerator is linear because
it is more difficult to reach higher energies due to synchrotron radiation
in the circular electron accelerators. In the reference \cite{SPPC-based},
four types of electron proton colliders have been recently proposed
by using the parameters of known linear electron collider projects,
namely the ILC and PWFALC (Plasma Wake Field Accelerator Linear Collider)
\cite{plasma wakefield}. 

In this paper the production potential of the excited electrons at
these four electron-proton colliders are invetigated using the parameters
in the table 2, in which two parameter options ($35.6$ from the Pre-CDR
and $68$ TeV) for the energy of the proton beam are used. It is expected
that the luminosity values given in this table can be increased at
later stages of the project. So, in the analysis part of this work
calculations were made using more than one luminosity value for each
collider. For such SPPC-based electron-proton colliders suitable detectors
have not been designed yet. Thus, this analysis does not include any
detector simulation and it is at the parton level.

\begin{table}
\caption{The main parameters of the SPPC-based lepton-hadron colliders.}

\begin{tabular}{|c|c|c|c|c|}
\hline 
Colliders  & $E_{e}$$\left(TeV\right)$  & $E_{p}$$\left(TeV\right)$  & $\sqrt{s}$ $\left(TeV\right)$  & $L_{int}$ $\left(cm^{-2}s^{-1}\right)$ \tabularnewline
\hline 
\hline 
ILC-SPPC1  & $0.5$  & $35.6$  & $8.44$  & $2.51$$\times$ $10^{31}$\tabularnewline
\hline 
ILC-SPPC2  & $0.5$  & $68$  & $11.66$  & $6.45$$\times$ $10^{31}$ \tabularnewline
\hline 
PWFALC-SPPC1  & $5$  & $35.6$  & $26.68$  & $7.37$$\times$ $10^{30}$ \tabularnewline
\hline 
PWFALC-SPPC2  & $5$  & $68$  & $36.88$  & $1.89$$\times$ $10^{31}$ \tabularnewline
\hline 
\end{tabular}
\end{table}

\section{effective lagrang\i an, decay w\i dths and cross-sect\i ons}

The gauge interaction of a spin-1/2 excited lepton with the ordinary
leptons and a gauge boson ($\gamma$,$Z,W^{\pm}$) is described by
following SU(2)xU(1) invariant Lagrangian \cite{excited lepton-LEP-HERA,excited quark and lepton,looking for,excited fermions},

\begin{center}
\begin{equation}
L=\frac{1}{2\Lambda}\bar{l_{R}^{\star}}\sigma^{\mu\nu}[fg\frac{\overrightarrow{\tau}}{2}.\overrightarrow{W}_{\mu\nu}+f'g'\frac{Y}{2}B_{\mu\nu}]l_{L}+h.c.,
\end{equation}

\par\end{center}

where $l$ and $l^{\star}$ denote ordinary lepton and the excited
lepton, respectively, $\Lambda$ is the compositeness scale, $\overrightarrow{W}_{\mu\nu}$
and $B_{\mu\nu}$ are the field strength tensors, $g$ and $g'$ are
the gauge couplings , $f$ and $f'$ are the scaling factors, Y is
hypercharge, $\sigma^{\mu\nu}=i(\gamma^{\mu}\gamma^{\nu}-\gamma^{\nu}\gamma^{\mu})/2$
where $\gamma^{\mu}$ are the Dirac matrices, and $\overrightarrow{\tau}$
represents the Pauli matrices.

For the excited electrons, three decay modes are possible: radiative
decay $e^{\star}\rightarrow e\gamma$, neutral weak decay $e^{\star}\rightarrow eZ$
and charged weak decay $e^{\star}\rightarrow\nu W^{-}$. Photon channel
is preferred in this study because it can be easily detected.

Neglecting the SM electron mass, the decay widths of the excited electrons
for the gauge interactions are given as, 

\begin{equation}
\varGamma(l^{\star}\rightarrow lV)=\frac{\alpha m^{\star3}}{4\Lambda^{2}}f_{V}^{2}(1-\frac{m_{V}^{2}}{m^{\star2}})^{2}(1+\frac{m_{V}^{2}}{2m^{\star2}}),
\end{equation}

where $m^{\star}$ is the mass of the excited electron, $f_{V}$ is
the new electroweak coupling parameter corresponding to the gauge
boson V, where V=$W,Z,\gamma,$ and $f_{\gamma}=-(f+f')/2$, $f_{Z}=(-f\cot\theta_{W}+f\tan\theta_{W})/2$,
$f_{W}=(f/\sqrt{2}\sin\theta_{W})$, where $\theta_{W}$ is the weak
mixing angle, and $m_{V}$ is the mass of the gauge boson.

\begin{figure}
\centering{}\includegraphics{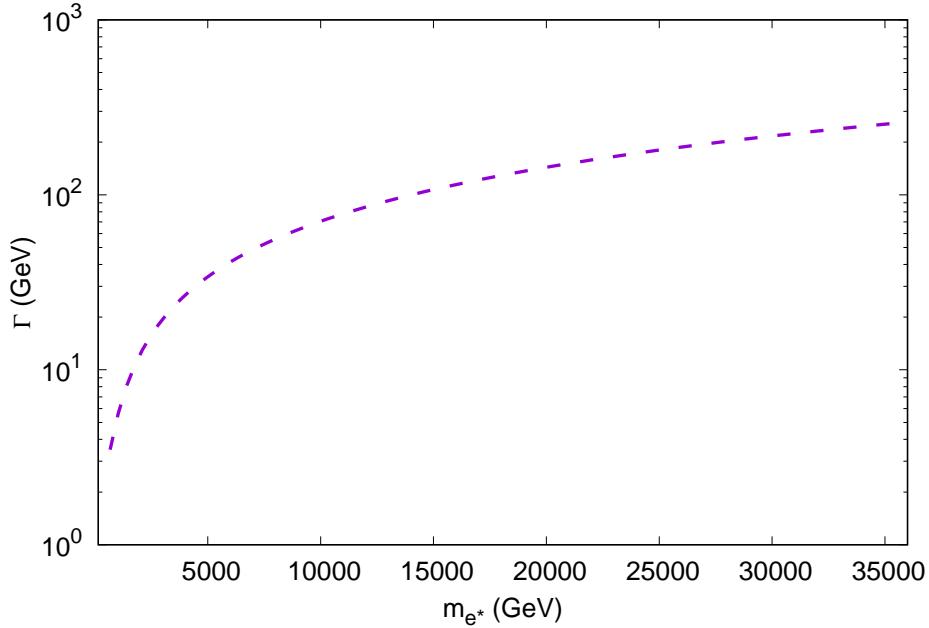}\caption{The total decay widths of the excited electron for the energy scale
$\varLambda=m_{e^{\star}}$}
\end{figure}

For the numerical calculations we have used the program CALCHEP \cite{calchep}
for signal event production and MADGRAPH \cite{automated computation}
for background event production, which are high-energy simulation
programmes. We present the total decay widths of the excited electron
in Figure 1 for the energy scale $\varLambda=m_{e^{\star}}$. Figure
2 shows the total cross-sections for the excited electron production
at the SPPC-based four electron-proton colliders, using the CTEQ6L1
parton distribution functions \cite{parton}.

\begin{figure}
\begin{centering}
\includegraphics{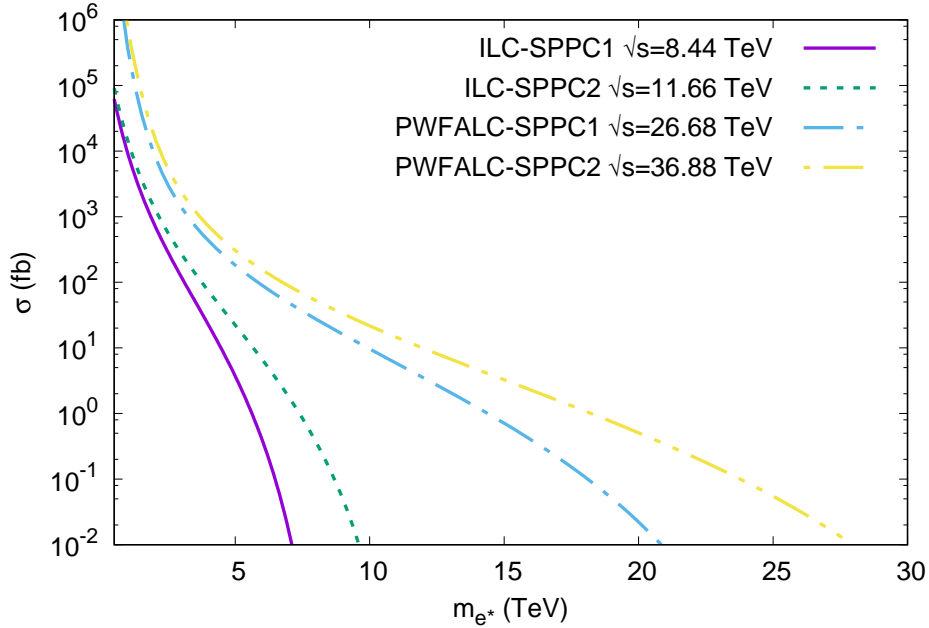}\caption{The total production cross-sections of the excited electrons with
respect to its mass at the SPPC-based electron-hadron colliders with
various center-of-mass energies for $\varLambda=m_{e^{\star}}$ and
the coupling $f=f'=1$.}

\par\end{centering}

\end{figure}

\section{SIGNAL AND BACKGROUND ANALYSIS}

We have analyzed in this section the potentials of the future lepton-hadron
colliders to search for the excited electrons through the single production
process $ep\rightarrow e^{\star}X$ with subsequent decays of the
excited electrons into an electron and photon. Thus, we consider the
process $ep\rightarrow e,\gamma,j$ for signal and background, where
$j$ represents jets that are composed of quarks ($u,\overline{u},d,\overline{d},c,\overline{c},s,\overline{s},b,\bar{b}$).
The subprocesses are $eq(\overline{q})\rightarrow e\gamma q(\overline{q})$,
where q represents quarks ($u,d,c,s,b)$ and $\bar{q}$ represents
anti-quarks ($\overline{u},\overline{d},\overline{c},\overline{s},\bar{b}$).
The Feynman diagrams for the signal process are shown in Fig.3. We
discuss here the differences in the signal and background over some
kinematical quantities of final-state particles to assign the kinematical
cuts best suited for the discovery of the excited electrons.

\begin{figure}
\begin{centering}
\includegraphics[scale=0.7]{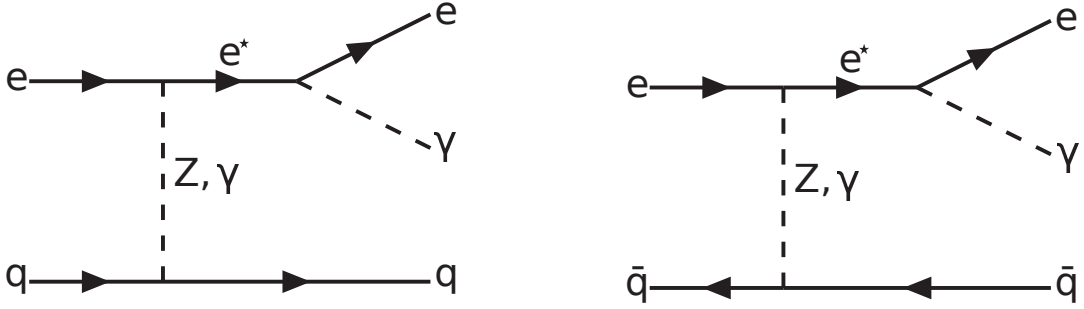}
\par\end{centering}

\caption{Leading-order Feynman diagrams for the signal process $ep\rightarrow e,\gamma,j$. }

\end{figure}

There are four electron-proton colliders with different center-of-mass
energies. The ILC parameters for the first two (called ILC-SPPC1 and
ILC-SPPC2) and the PWFALC parameters for the last two (called PWFALC-SPPC1
and PWFALC-SPPC2) are used according to the Table 2. Thus, in the
following sub-sections these colliders are covered under two categories
as ILC-SPPC and PWFALC-SPPC colliders.

\subsection*{ILC-SPPC Colliders}

In order to separate the excited electron signals from the background
we have firstly applied pre-selection cuts to the transverse momentum
of the electron, photon and jets in the final-state as $P_{T}^{e,\gamma,j}>20$
GeV. We have generated $10^{5}$ events for both the signal and background,
and plotted angular distributions and tranverse momentum distributions
of the final-state particles using these event files. The event generation
for the signal was made for various mass values at the specific intervals
starting from $3000$ GeV. Because both ILC-SPPC colliders have similar
distributions, here's just one of them shown.

\begin{figure}
\begin{centering}
\includegraphics[scale=0.45]{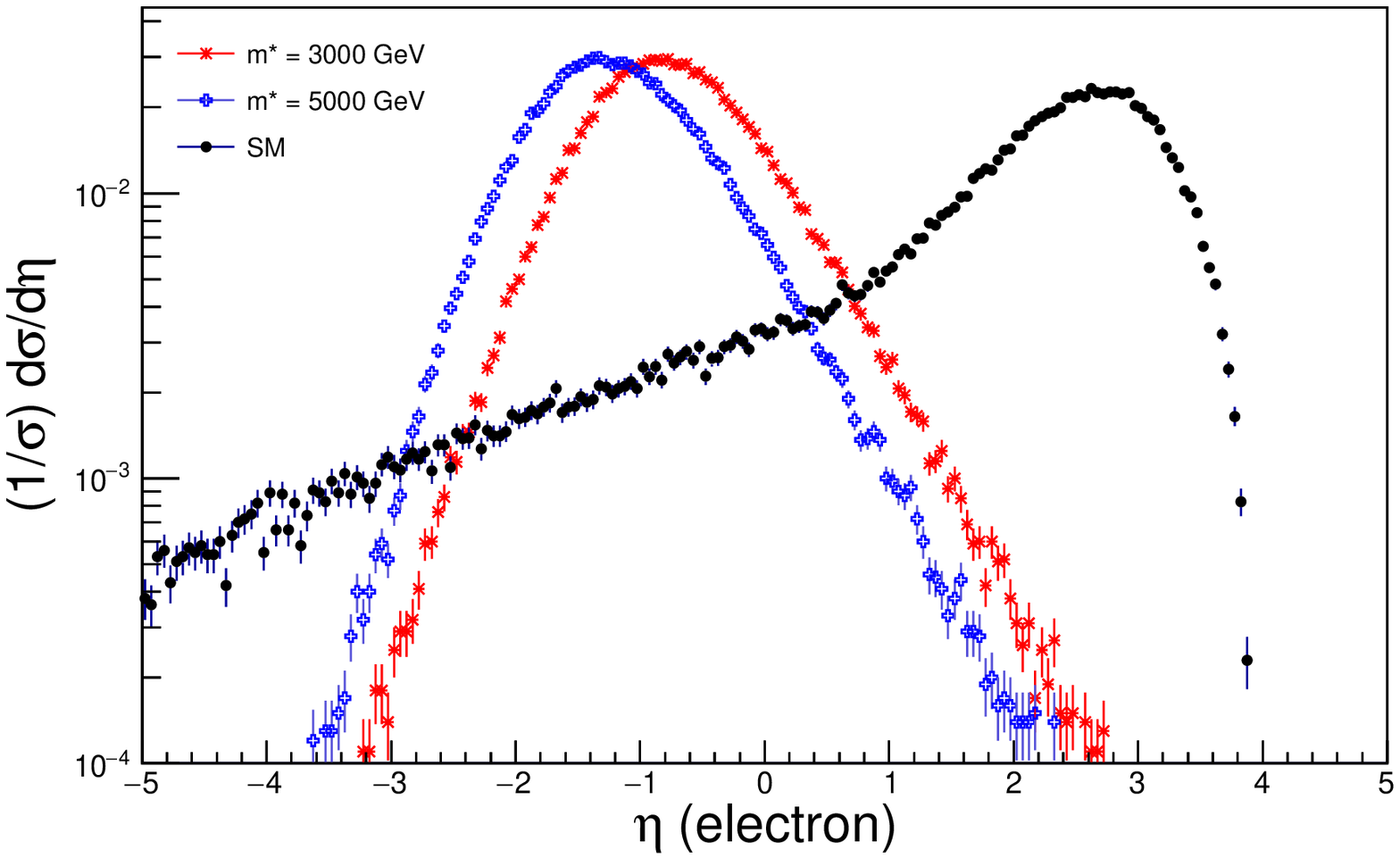}\includegraphics[scale=0.45]{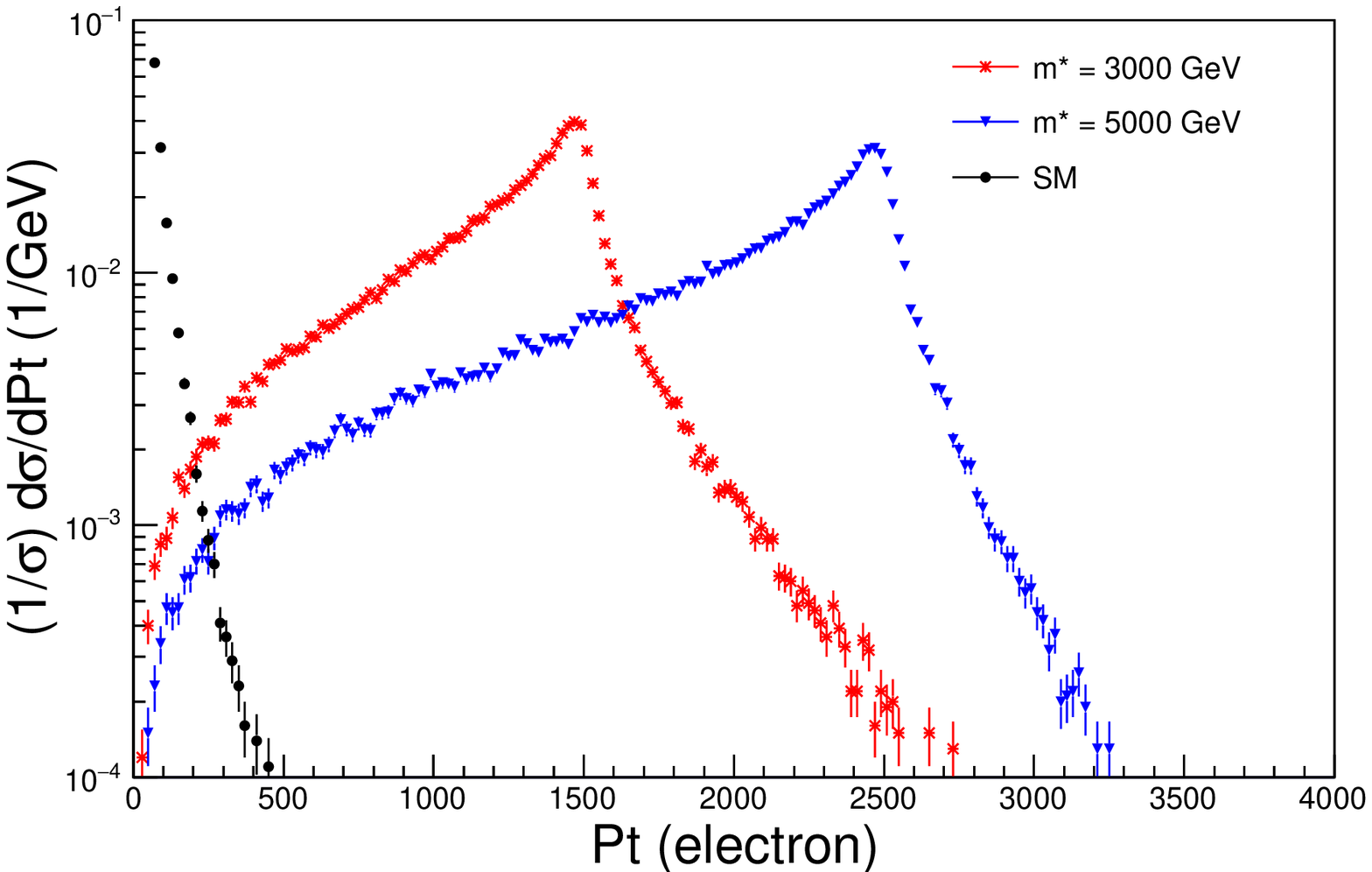}
\par\end{centering}

\begin{centering}
\includegraphics[scale=0.45]{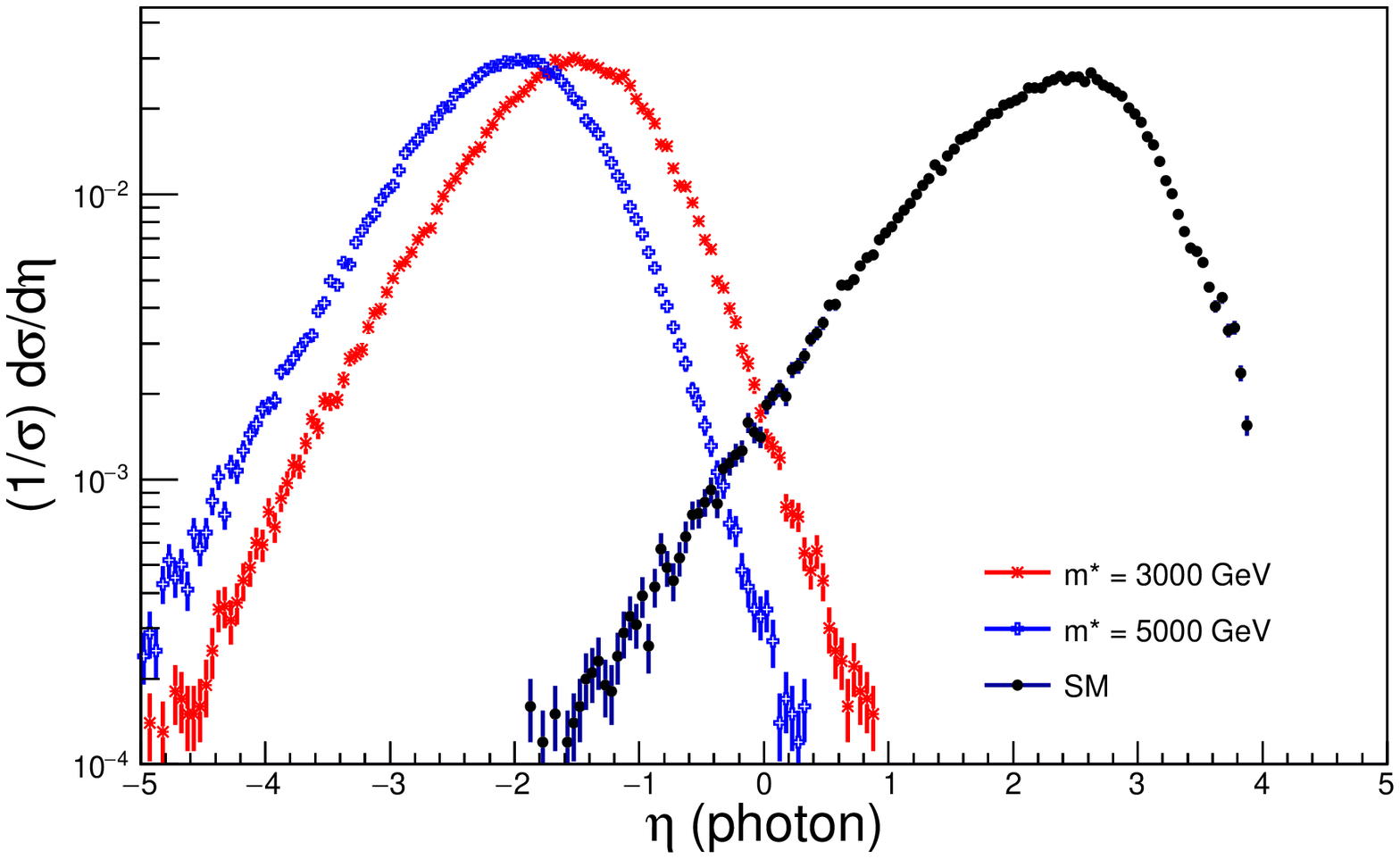}\includegraphics[scale=0.45]{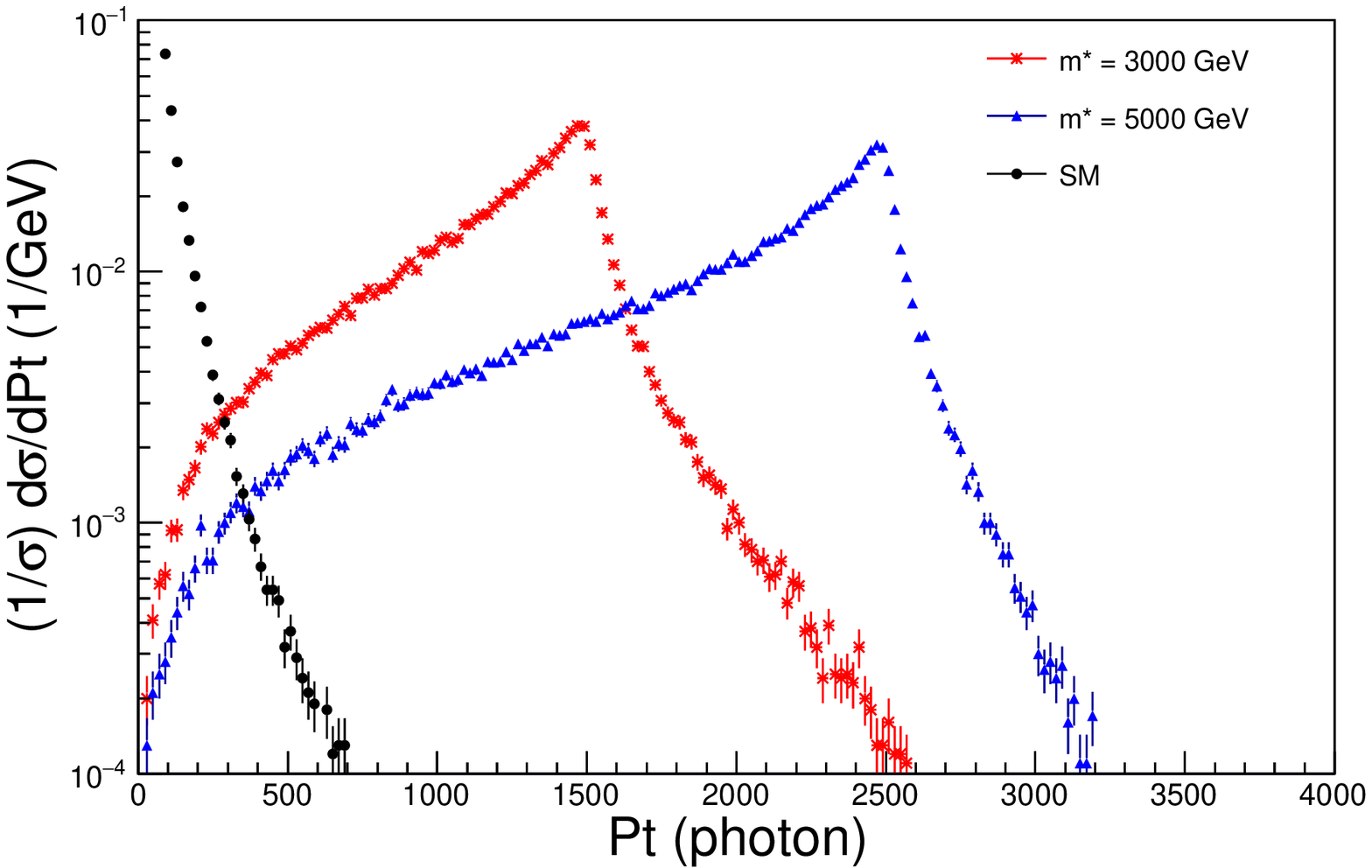}
\par\end{centering}

\caption{Some kinematical distributions of particles in the final-state for
the ILC-SPPC collider with $\sqrt{s}=8.4$ TeV. In the left-top and
right-top panels we show normalized pseudorapidity and transverse
momentum distributions of the electron, respectively. The similar
distributions for photon are shown in the lower panels. All distributions
contain statistical errors.}

\end{figure}

Figure 4 shows angular distributions and transverse momentum distributions
of the electron and photon in the final-state for the ILC-SPPC collider
with $\sqrt{s}=8.4$ TeV. The pseudorapidity distributions (top-left
and bottom-left) of the signal are peaked almost at $\eta=[-1,-2]$
interval for given parameter values ($m_{e^{\star}}=3000,$ $5000$
GeV and $\varLambda=m_{e^{\star}}$) of both particles. Since the
pseudorapidity is defined as $\eta=-\ln\tan(\theta/2)$, where $\theta$
is polar angle, the electrons and photons are of backward, as a result
the excited electrons are produced in the backward direction. Also,
the separation of the signal and background from each other is very
well for all mass values. To reduce the contributions from the background
we applied a cut on the pseudorapidity for both particles as $-3<\eta^{e}<0.4$
and $-5<\eta^{\gamma}<0$.

When we look at the transverse momentum distributions in the top-right
and bottom-right panels of the same figure, it is seen that the signal
and background are still separated from each other for both particles.
Thus, we applied a cut at $300$ GeV (for electron) and $400$ GeV
(for photon) to reduce the background. These cuts do not affect the
signal too much.

As for the ILC-SPPC collider with $\sqrt{s}=11.6$ TeV (ILC-SPPC2),
similar distributions in large part have been obtained. Similar analyzes
have been made and the kinematical cuts best suited for discovery
of the excited electrons have been determined as $-3<\eta^{e}<1$,
$-5<\eta^{\gamma}<0.2$, $p_{T}^{e}>300$, $p_{T}^{\gamma}>400$ GeV.

We calculated the selection efficiencies to see how the applied discovery
cuts affected the signal and the background. As can be seen from Table
3 and 4, the efficiency values for the signal are greater than 90\%
for both ILC-SPPC colliders. This shows that the discovery cuts applied
have very little effect on the signal. As for the background, the
efficiency values are close to zero because the background is drastically
reduced. These tables also show the cross-sections before and after
the discovery cuts for both colliders.

\begin{table}
\caption{The efficiencies of the standard model background and of our signature
after the application of all discovery cuts for the ILC-SPPC collider
with $\sqrt{s}=8.4$ TeV.}

\begin{tabular}{|c|c|c|c|c|}
\hline 
\multicolumn{2}{|c|}{} & $\sigma$ before cut (fb) & $\sigma$ after cut (fb) & efficiency ($\epsilon$)\tabularnewline
\hline 
\hline 
\multicolumn{2}{|c|}{background} & $22100$ & $10.608$ & $0.00048$\tabularnewline
\hline 
\multirow{11}{*}{$m_{e^{\star}}$(GeV) } & $3000$ & $12.8$ & $11.5$ & $0.89$\tabularnewline
\cline{2-5} 
 & $3500$ & $5.99$ & $5.54$ & $0.92$\tabularnewline
\cline{2-5} 
 & $4000$ & $2.78$ & $2.61$ & $0.93$\tabularnewline
\cline{2-5} 
 & $4500$ & $1.24$ & $1.18$ & $0.95$\tabularnewline
\cline{2-5} 
 & $5000$ & $0.52$ & $0.50$ & $0.96$\tabularnewline
\cline{2-5} 
 & $5500$ & $0.19$ & $0.18$ & $0.94$\tabularnewline
\cline{2-5} 
 & $6000$ & $0.064$ & $0.062$ & $0.96$\tabularnewline
\cline{2-5} 
 & $6500$ & $0.017$ & $0.016$ & $0.94$\tabularnewline
\cline{2-5} 
 & $7000$ & $0.0033$ & $0.0032$ & $0.96$\tabularnewline
\cline{2-5} 
 & $7500$ & $0.00047$ & $0.00045$ & $0.95$\tabularnewline
\cline{2-5} 
 & $8000$ & $0.000107$ & $0.000101$ & $0.94$\tabularnewline
\hline 
\end{tabular}
\end{table}

\begin{table}
\caption{The efficiencies of the standard model background and of our signature
after the application of all discovery cuts for the ILC-SPPC collider
with $\sqrt{s}=11.6$ TeV.}

\begin{tabular}{|c|c|c|c|c|}
\hline 
\multicolumn{2}{|c|}{} & $\sigma$ before cut (fb) & $\sigma$ after cut (fb) & efficiency ($\epsilon$)\tabularnewline
\hline 
\hline 
\multicolumn{2}{|c|}{background} & $29600$ & $15.98$ & $0.00053$\tabularnewline
\hline 
\multirow{15}{*}{$m_{e^{\star}}$(GeV) } & $3000$ & $27.5$ & $26.04$ & $0.94$\tabularnewline
\cline{2-5} 
 & $3500$ & $15.1$ & $14.51$ & $0.96$\tabularnewline
\cline{2-5} 
 & $4000$ & $8.53$ & $8.27$ & $0.96$\tabularnewline
\cline{2-5} 
 & $4500$ & $4.94$ & $4.82$ & $0.97$\tabularnewline
\cline{2-5} 
 & $5000$ & $2.89$ & $2.82$ & $0.97$\tabularnewline
\cline{2-5} 
 & $5500$ & $1.62$ & $1.58$ & $0.97$\tabularnewline
\cline{2-5} 
 & $6000$ & $0.937$ & $0.917$ & $0.97$\tabularnewline
\cline{2-5} 
 & $6500$ & $0.499$ & $0.489$ & $0.97$\tabularnewline
\cline{2-5} 
 & $7000$ & $0.262$ & $0.256$ & $0.97$\tabularnewline
\cline{2-5} 
 & $7500$ & $0.129$ & $0.125$ & $0.96$\tabularnewline
\cline{2-5} 
 & $8000$ & $0.059$ & $0.058$ & $0.98$\tabularnewline
\cline{2-5} 
 & $8500$ & $0.025$ & $0.024$ & $0.96$\tabularnewline
\cline{2-5} 
 & $9000$ & $0.0093$ & $0.0089$ & $0.95$\tabularnewline
\cline{2-5} 
 & $9500$ & $0.0029$ & $0.0028$ & $0.96$\tabularnewline
\cline{2-5} 
 & $10000$ & $0.00078$ & $0.00074$ & $0.94$\tabularnewline
\hline 
\end{tabular}
\end{table}

Another way to see the effect of the discovery cuts applied is to
draw invariant mass distributions for both the signal and background.
Figure 5 shows the invariant mass distributions of the excited electron
and the corresponding background after the application of all discovery
cuts for the ILC-SPPC colliders. It is clearly seen from this figure
that the background curves are below the signal peaks for the both
colliders.

\begin{figure}
\begin{centering}
\includegraphics[scale=0.6]{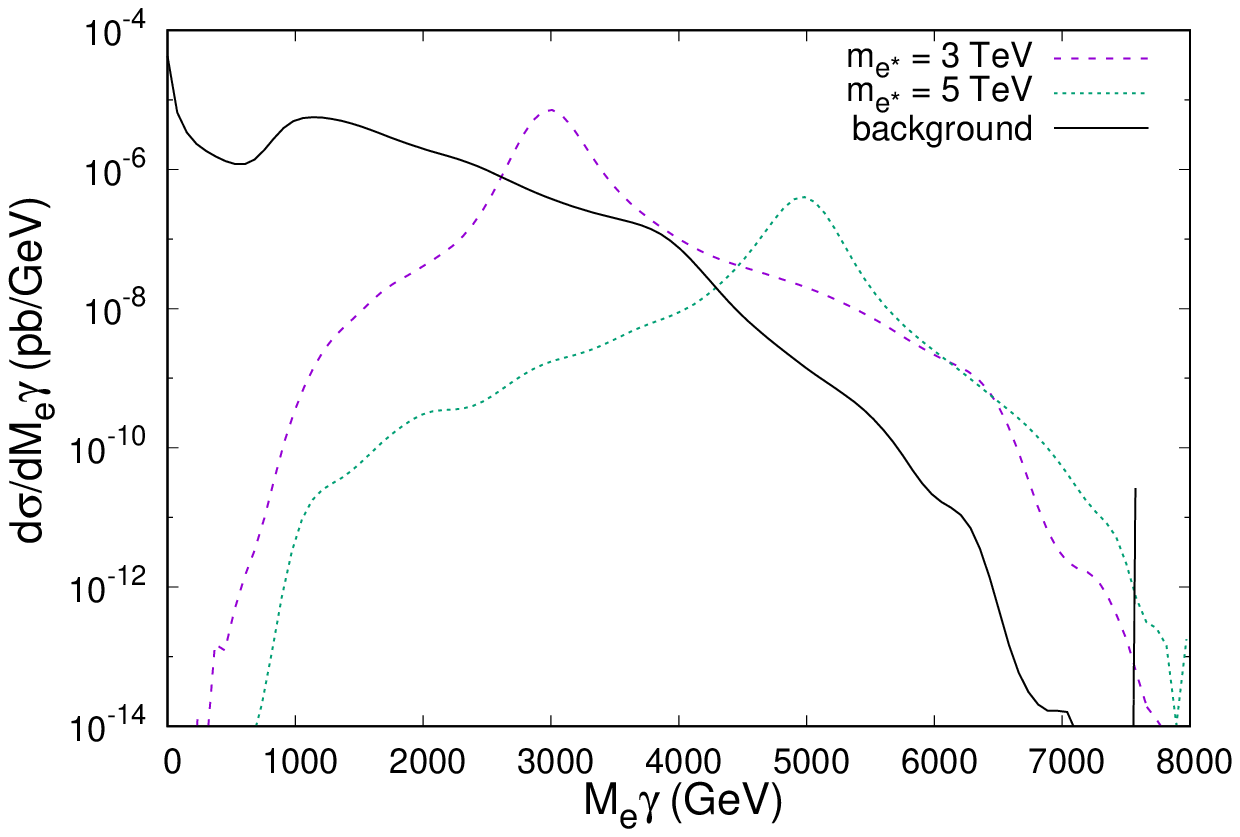}\includegraphics[scale=0.6]{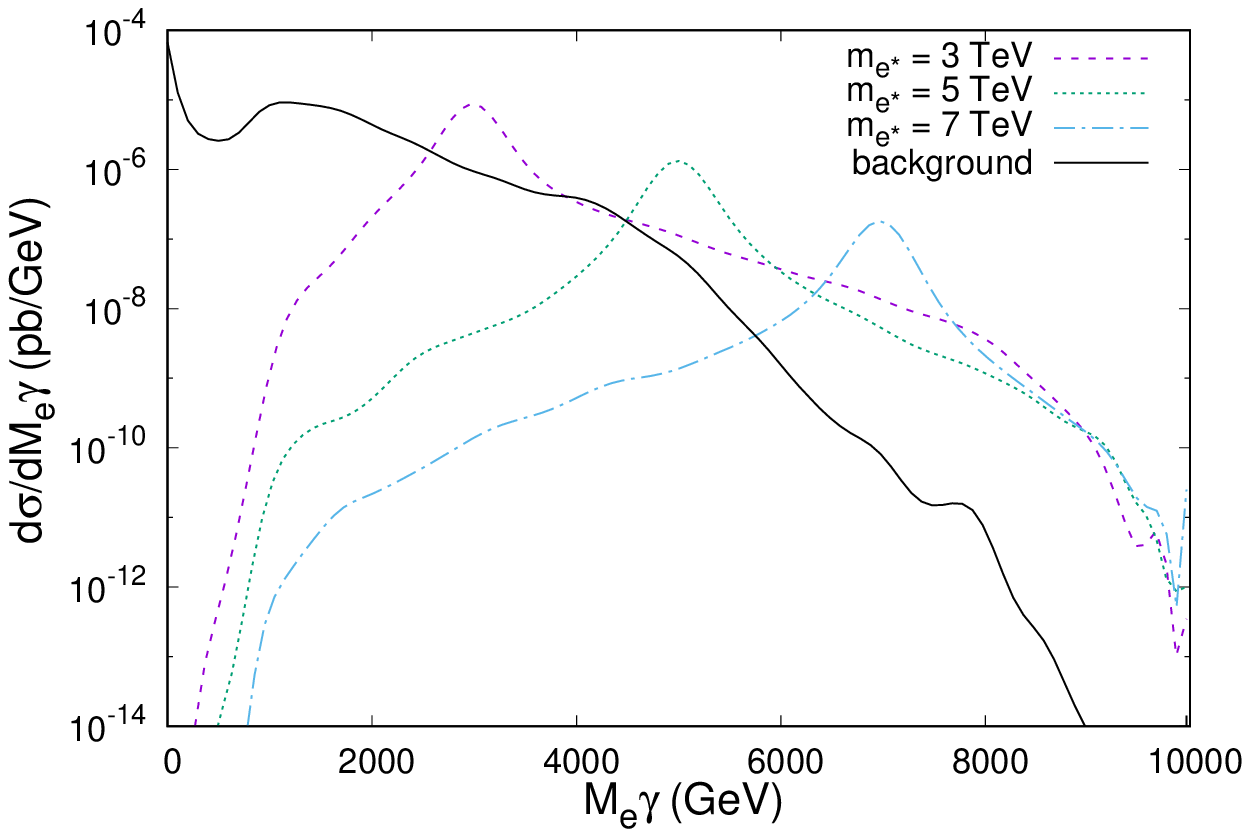}
\par\end{centering}

\caption{The distributions in the invariant mass of the excited electron signal
and the corresponding background for the ILC-SPPC colliders. The left
panel shows the distributions at the collider with $\sqrt{s}=8.4$
TeV for the excited electron masses of $3$ and $5$ TeV, and the
right panel shows the distributions at the collider with $\sqrt{s}=11.6$
TeV for the excited electron masses of $3$,$5$ and $7$ TeV.}

\end{figure}

In order to calculate the statistical significance (SS) of the expected
signal yield, we have used the formula of

\begin{equation}
SS=\sqrt{2\left[\left(S+B\right)\ln\left(1+\left(\frac{S}{B}\right)\right)-S\right]},
\end{equation}

where S and B denote event numbers of the signal and background, respectively.
In Fig. 6 we show the contour plots of $S=2$ ($2\sigma$-exclusion),
$3$ ($3\sigma$-observation) and $5$ ($5\sigma$-discovery) in the
parameter space ($L$, $m^{\star}$), where $L$ is integrated luminosity
of the collider, for both ILC-SPPC colliders. We consider a scan of
the parameter space ($L$, $m^{\star}$) within the range L $\in$
$[1,100]$ $fb^{-1}$ with step of $10$ $fb^{-1}$. In these distributions
the regions below the $S=5$ curve are excluded.

\begin{figure}
\includegraphics[scale=0.45]{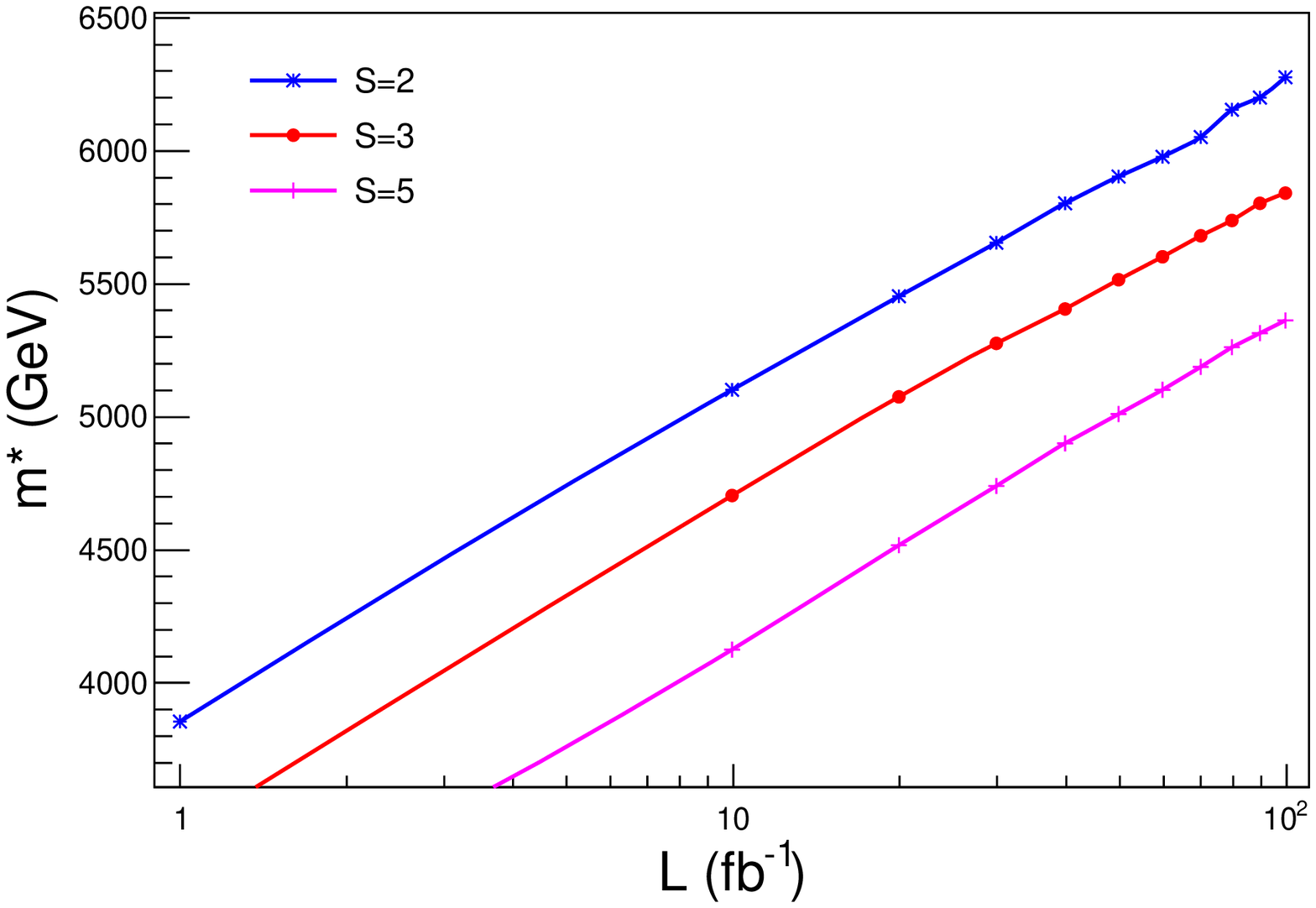}\includegraphics[scale=0.45]{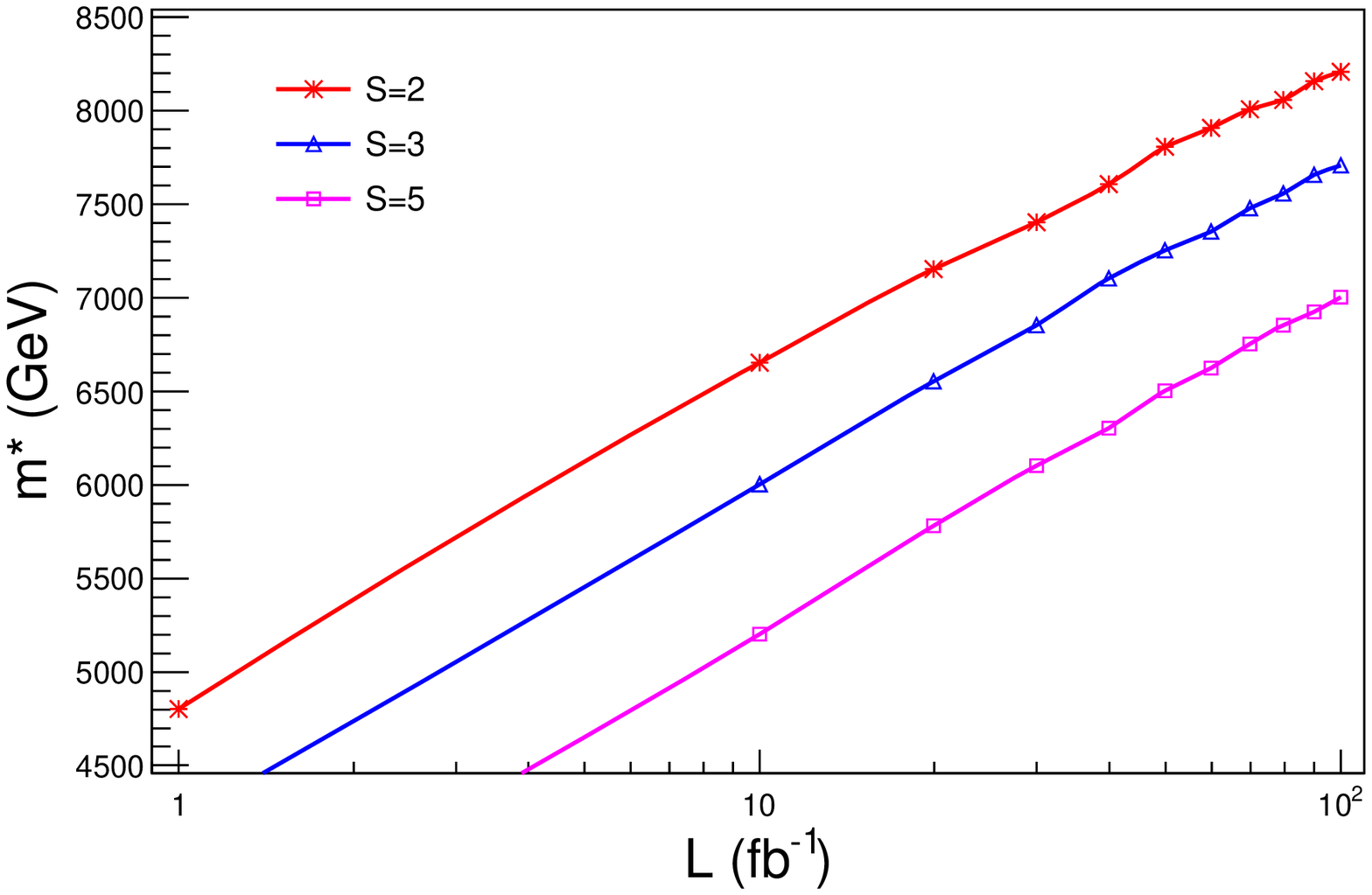}

\caption{Contour maps of the statistical significance for S=$2$, $3$ and
$5$ in the parameter space (L, $m^{\star}$) for the ILC-SPPC collider
with $\sqrt{s}=8.4$ (left) and $11.6$ TeV (right).}

\end{figure}

\subsection*{PWFALC-SPPC Colliders}

If the excited electrons had not been observed at the ILC-SPPC colliders,
they would have been explored up to the mass of $26.6$ TeV and $36.8$
TeV at the PWFALC-SPPC1 and PWFALC-SPPC2 colliders, respectively.
To separate the signal from the background we have required kinematical
cuts $P_{T}^{e,\gamma,j}>20$ GeV, as for the ILC-SPPC colliders.
Figure 7 shows angular distributions and transverse momentum distributions
of the electron and photon in the final-state for the PWFALC-SPPC
collider with $\sqrt{s}=26.6$ TeV. It has been seen that the pseudorapidity
distributions (top-left and bottom-left) with low mass values like
$3000$ GeV of the signal are in the positive region compared to those
of the ILC-SPPC colliders. Therefore, some of the excited electrons
in these colliders are produced in the forward direction. On the other
hand, the signal and background curves are well separated from each
other for all mass values. To greatly reduce the effect of the background
we applied a cut on these distributions for both particles as $-2<\eta^{e}<3.2$
and $-5<\eta^{\gamma}<2.4$. The pseudorapidity cuts for the PWFALC-SPPC
collider with $\sqrt{s}=36.8$ TeV that has similar distributions
are $-2<\eta^{e}<3$ and $-5<\eta^{\gamma}<2.5$.

\begin{figure}
\includegraphics[scale=0.45]{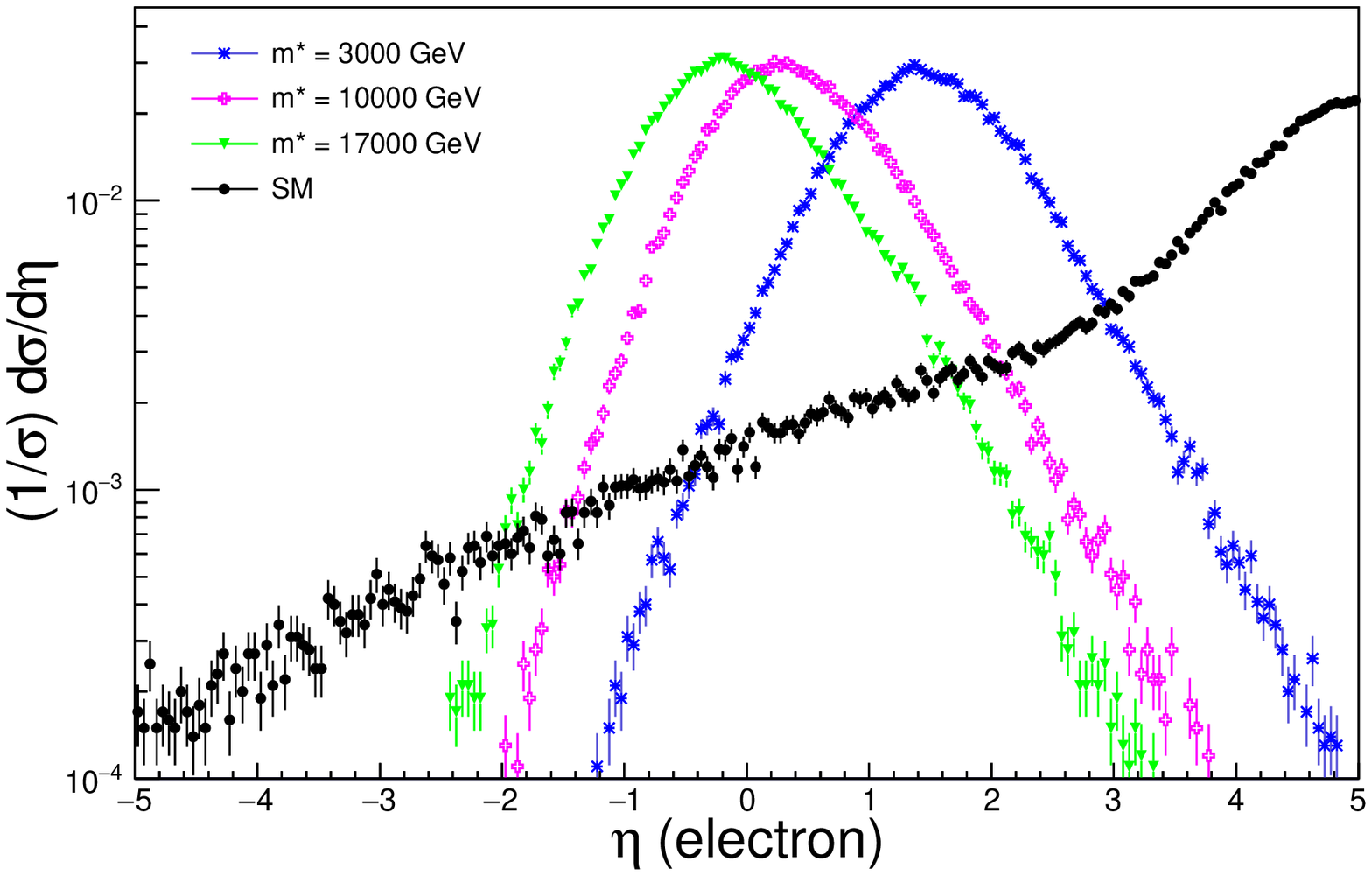}\includegraphics[scale=0.45]{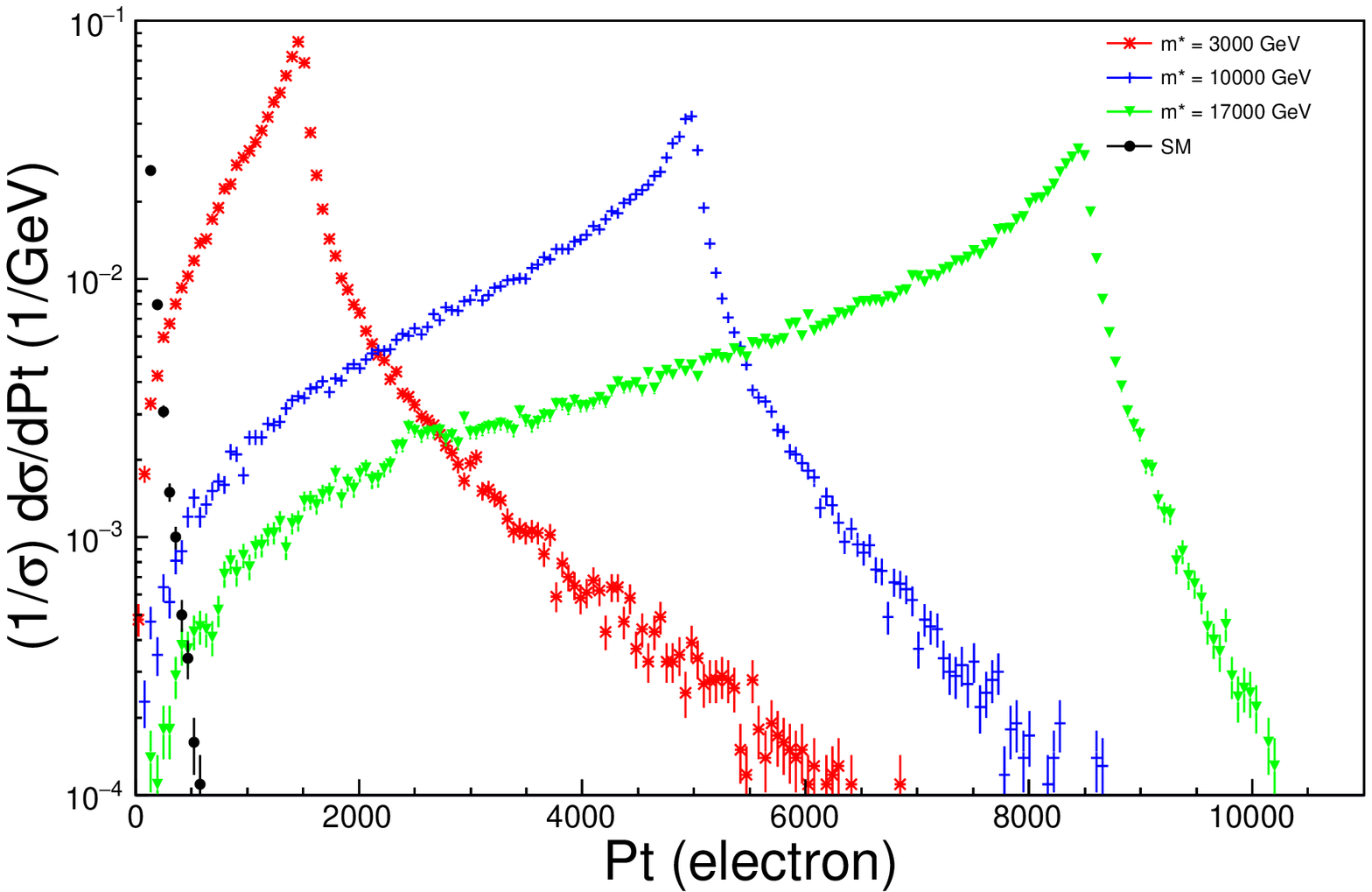}

\includegraphics[scale=0.45]{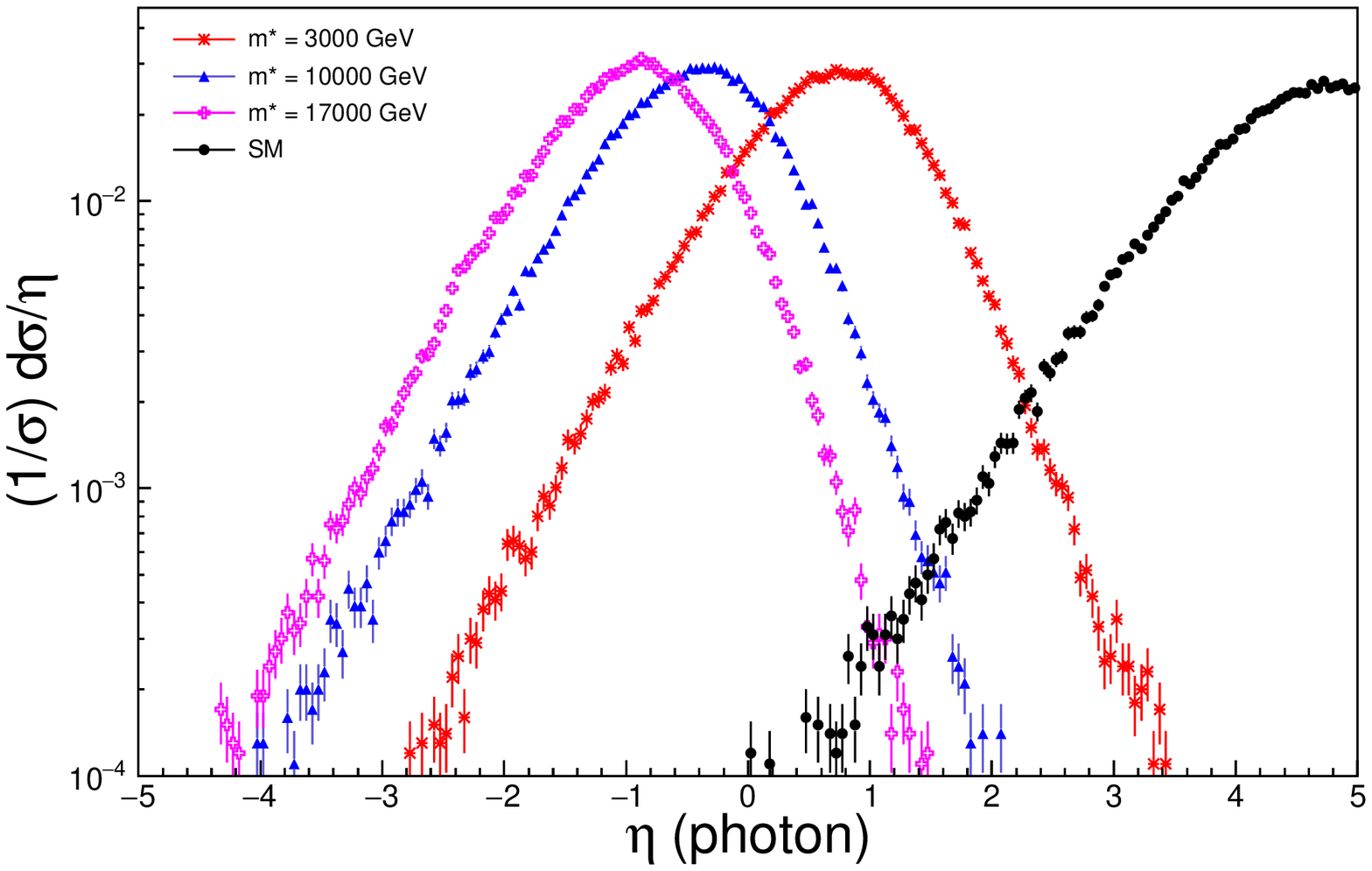}\includegraphics[scale=0.45]{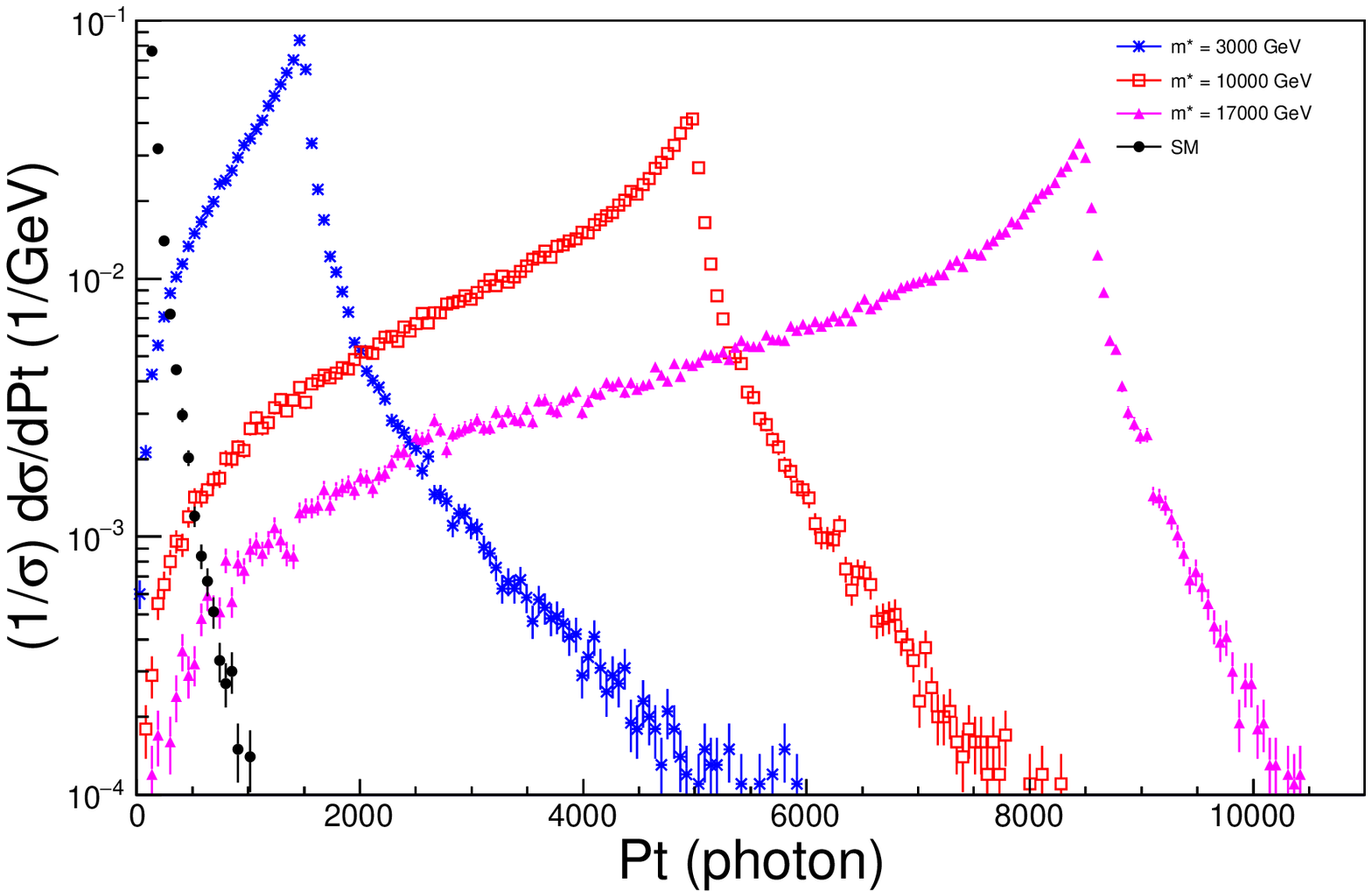}

\caption{Some kinematical distributions of particles in the final-state for
the PWFALC-SPPC collider with $\sqrt{s}=26.6$ TeV. In the left-top
and right-top panels we show normalized pseudorapidity and transverse
momentum distributions of the electron, respectively. The similar
distributions for photon are shown in the lower panels. All distributions
contain statistical errors.}

\end{figure}

In the transverse momentum distributions (top-right and bottom-right)
it has been seen that the signal and the background are again separated
from each other in a good way. The transverse momentum cuts for these
distributions are determined as $p_{T}^{e}>400$, $p_{T}^{\gamma}>500$
GeV for the both colliders. Table 5 and 6 show the selection efficiencies
after the application of all kinematical cuts for both colliders.
These tables show that the applied cuts drastically reduce the background
and have little effect on the signal.

\begin{table}
\caption{The efficiencies of the standard model background and of our signature
after the application of all discovery cuts for the PWFALC-SPPC collider
with $\sqrt{s}=26.6$ TeV.}

\begin{tabular}{|c|c|c|c|c|}
\hline 
\multicolumn{2}{|c|}{} & $\sigma$ before cut (fb) & $\sigma$ after cut (fb) & efficiency ($\epsilon$)\tabularnewline
\hline 
\hline 
\multicolumn{2}{|c|}{background} & $60100$ & $15.62$ & $0.00025$\tabularnewline
\hline 
\multirow{20}{*}{$m_{e^{\star}}$(GeV) } & $3000$ & $108$ & $98.9$ & $0.91$\tabularnewline
\cline{2-5} 
 & $4000$ & $42$ & $40.1$ & $0.95$\tabularnewline
\cline{2-5} 
 & $5000$ & $20$ & $19.4$ & $0.97$\tabularnewline
\cline{2-5} 
 & $6000$ & $10.5$ & $10.3$ & $0.98$\tabularnewline
\cline{2-5} 
 & $7000$ & $5.9$ & $5.8$ & $0.98$\tabularnewline
\cline{2-5} 
 & $8000$ & $3.4$ & $3.3$ & $0.97$\tabularnewline
\cline{2-5} 
 & $9000$ & $2$ & $1.9$ & $0.95$\tabularnewline
\cline{2-5} 
 & $10000$ & $1.2$ & $1.19$ & $0.99$\tabularnewline
\cline{2-5} 
 & $11000$ & $0.792$ & $0.786$ & $0.99$\tabularnewline
\cline{2-5} 
 & $12000$ & $0.491$ & $0.488$ & $0.99$\tabularnewline
\cline{2-5} 
 & $13000$ & $0.299$ & $0.297$ & $0.99$\tabularnewline
\cline{2-5} 
 & $14000$ & $0.178$ & $0.177$ & $0.99$\tabularnewline
\cline{2-5} 
 & $15000$ & $0.104$ & $0.103$ & $0.99$\tabularnewline
\cline{2-5} 
 & $16000$ & $0.0586$ & $0.0583$ & $0.99$\tabularnewline
\cline{2-5} 
 & $17000$ & $0.032$ & $0.031$ & $0.96$\tabularnewline
\cline{2-5} 
 & $18000$ & $0.0163$ & $0.0162$ & $0.99$\tabularnewline
\cline{2-5} 
 & $19000$ & $0.00789$ & $0.00784$ & $0.99$\tabularnewline
\cline{2-5} 
 & $20000$ & $0.00352$ & $0.00349$ & $0.99$\tabularnewline
\cline{2-5} 
 & $21000$ & $0.00139$ & $0.00138$ & $0.99$\tabularnewline
\cline{2-5} 
 & $22000$ & $0.000489$ & $0.000485$ & $0.99$\tabularnewline
\hline 
\end{tabular}
\end{table}

\begin{table}
\caption{The efficiencies of the standard model background and of our signature
after the application of all discovery cuts for the PWFALC-SPPC collider
with $\sqrt{s}=36.8$ TeV.}

\begin{tabular}{|c|c|c|c|c|}
\hline 
\multicolumn{2}{|c|}{} & $\sigma$ before cut (fb) & $\sigma$ after cut (fb) & efficiency ($\epsilon$)\tabularnewline
\hline 
\hline 
\multicolumn{2}{|c|}{background} & $78400$ & $40.76$ & $0.00051$\tabularnewline
\hline 
\multirow{15}{*}{$m_{e^{\star}}$(GeV) } & $3000$ & $172.6$ & $158.2$ & $0.91$\tabularnewline
\cline{2-5} 
 & $5000$ & $33.2$ & $32.2$ & $0.96$\tabularnewline
\cline{2-5} 
 & $7000$ & $10.6$ & $10.4$ & $0.98$\tabularnewline
\cline{2-5} 
 & $9000$ & $4.33$ & $4.28$ & $0.98$\tabularnewline
\cline{2-5} 
 & $11000$ & $1.97$ & $1.95$ & $0.98$\tabularnewline
\cline{2-5} 
 & $13000$ & $0.959$ & $0.952$ & $0.99$\tabularnewline
\cline{2-5} 
 & $15000$ & $0.473$ & $0.470$ & $0.99$\tabularnewline
\cline{2-5} 
 & $17000$ & $0.2309$ & $0.2294$ & $0.99$\tabularnewline
\cline{2-5} 
 & $19000$ & $0.1109$ & $0.1100$ & $0.99$\tabularnewline
\cline{2-5} 
 & $21000$ & $0.0515$ & $0.0510$ & $0.99$\tabularnewline
\cline{2-5} 
 & $23000$ & $0.0221$ & $0.0218$ & $0.98$\tabularnewline
\cline{2-5} 
 & $25000$ & $0.00842$ & $0.00831$ & $0.98$\tabularnewline
\cline{2-5} 
 & $27000$ & $0.00282$ & $0.00277$ & $0.98$\tabularnewline
\cline{2-5} 
 & $29000$ & $0.000778$ & $0.000763$ & $0.98$\tabularnewline
\cline{2-5} 
 & $31000$ & $0.000163$ & $0.000159$ & $0.97$\tabularnewline
\hline 
\end{tabular}
\end{table}

When we examine the invariant mass distributions, shown in Fig.8 for
both colliders, obtained after applying the cuts it has been seen
that the signal peaks are above the background curve. 

\begin{figure}
\begin{centering}
\includegraphics[scale=0.45]{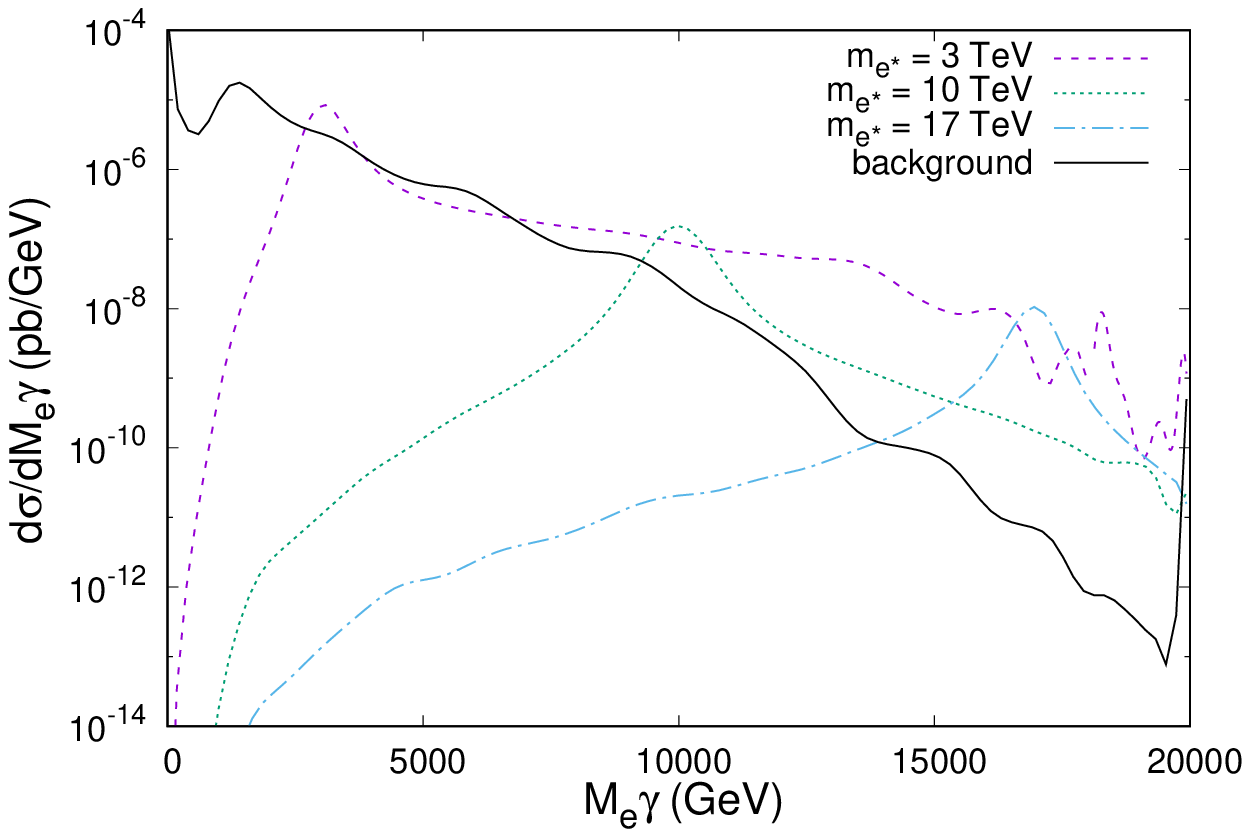}\includegraphics[scale=0.45]{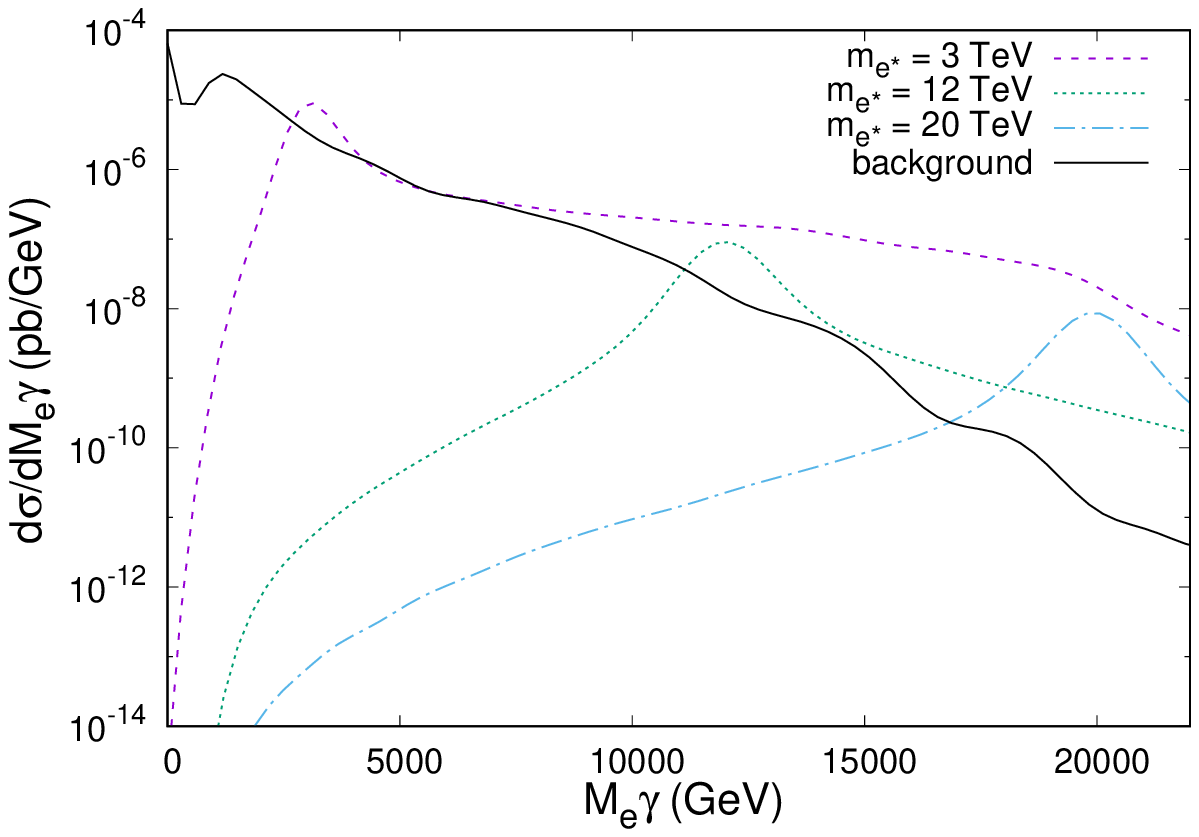}
\par\end{centering}

\caption{The distributions in the invariant mass of the excited electron signal
and the corresponding background for the PWFALC-SPPC colliders. The
left panel shows the distributions at the collider with $\sqrt{s}=26.6$
TeV for the excited electron masses of $3$, $10$ and $17$ TeV,
and the right panel shows the distributions at the collider with $\sqrt{s}=36.8$
TeV for the excited electron masses of $3$,$12$ and $20$ TeV.}

\end{figure}

And finally we plot the statistical significance curves of the expected
signal in the parameter space ($L$, $m^{\star}$) as shown in Fig.
9 for both colliders. The areas under the curve $S=5$ are the mass
values excluded by the collider.

\begin{figure}
\includegraphics[scale=0.45]{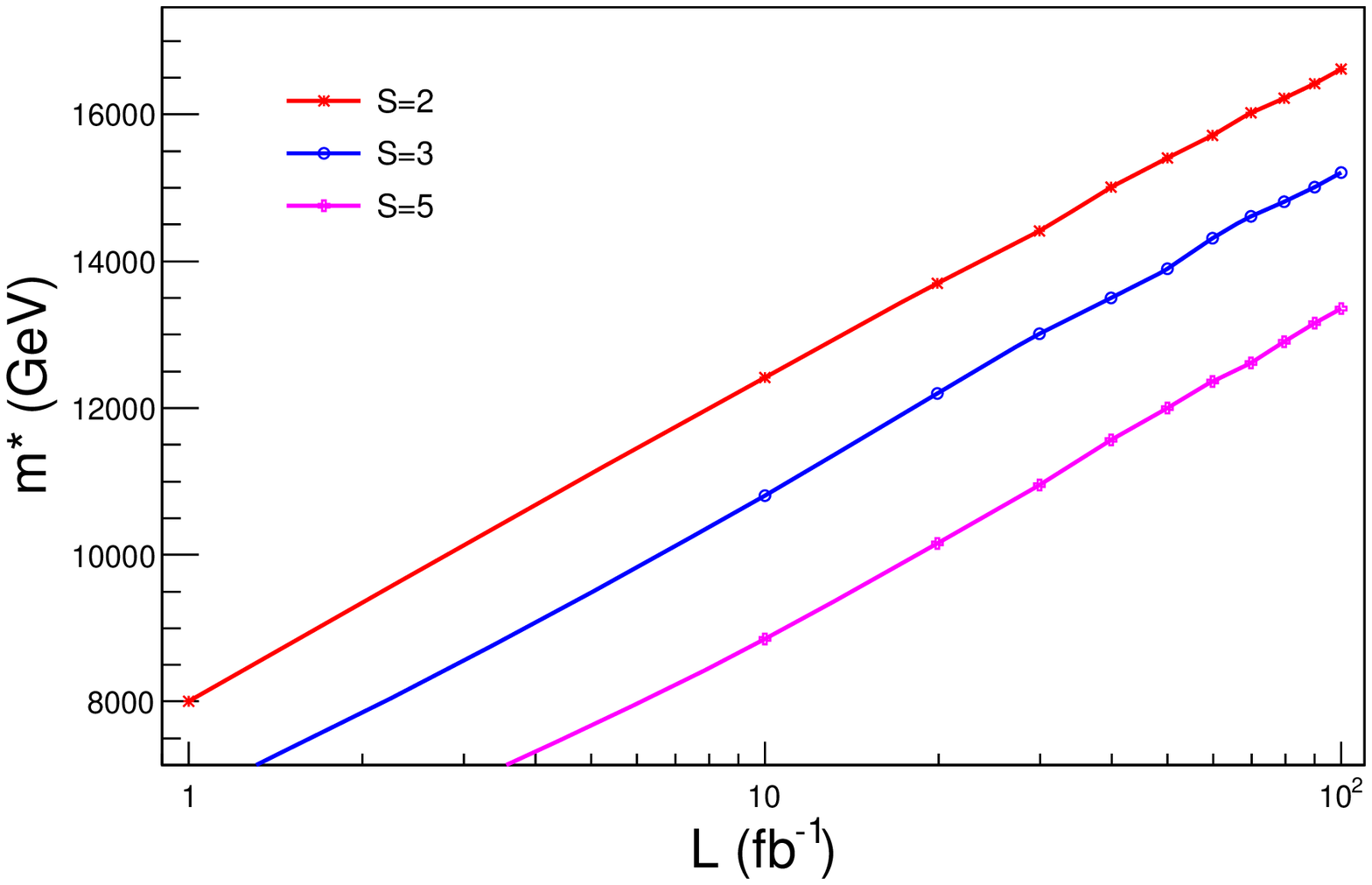}\includegraphics[scale=0.45]{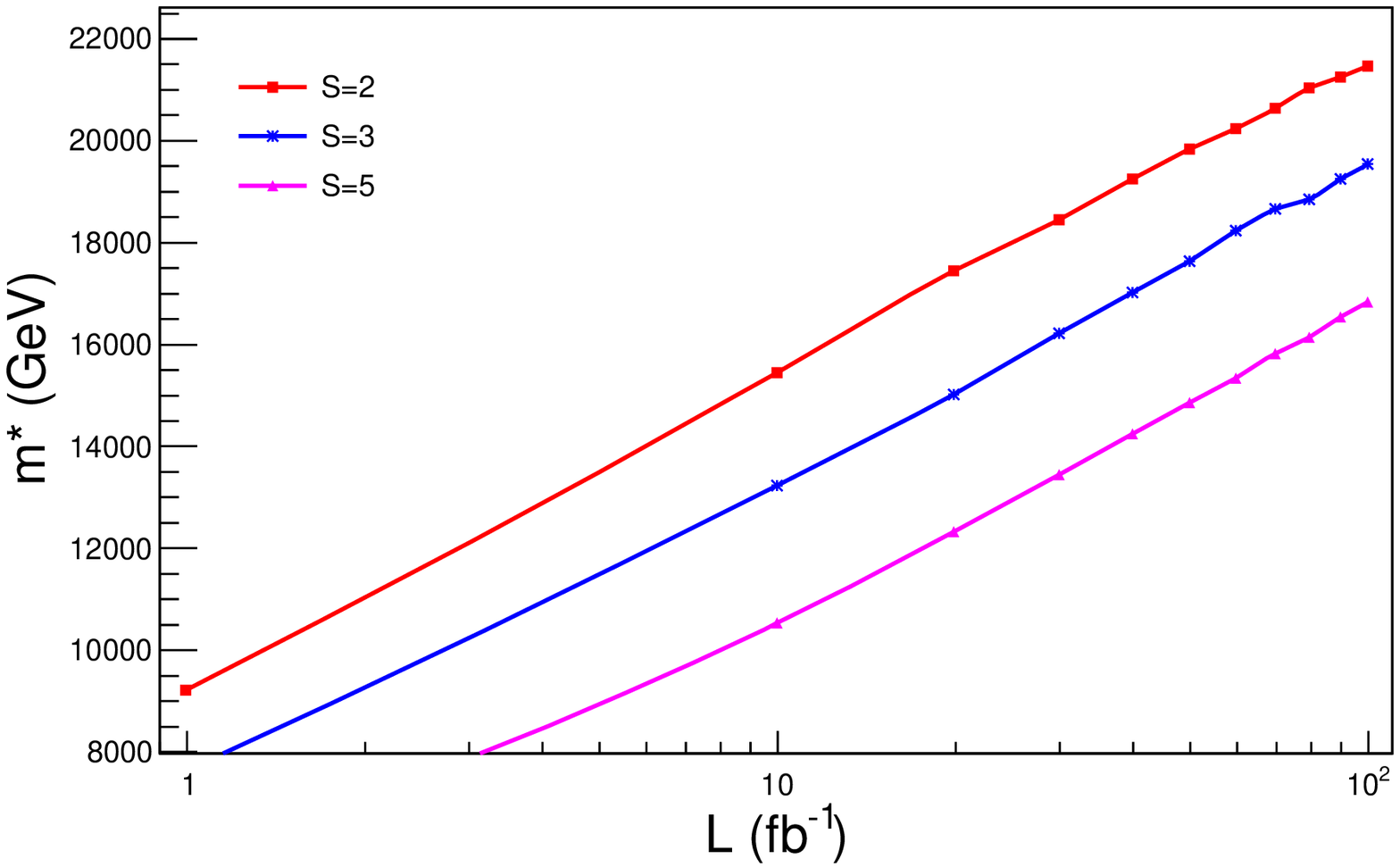}

\caption{Contour maps of the statistical significance for S=$2$, $3$ and
$5$ in the parameter space (L, $m^{\star}$) for the PWFALC-SPPC
collider with $\sqrt{s}=26.6$ (left) and $36.8$ TeV (right).}

\end{figure}

\section{CONCLUSION}

In this work, the production potential of the excited electrons predicted
by the composite models at the four SPPC-based lepton-hadron colliders,
namely ILC-SPPC1 ($\sqrt{s}=8.44$ TeV), ILC-SPPC2 ($\sqrt{s}=11.66$
TeV), PWFALC-SPPC1 ($\sqrt{s}=26.68$ TeV) and PWFALC-SPPC2 ($\sqrt{s}=36.88$
TeV) has been searched. In the all simulations it is assumed that
the energy scale is $\varLambda=m_{e^{\star}}$ and the coupling parameter
is $f=f'=1$. In the analysis, kinematical cuts required for the discovery
of the excited electrons were determined for the paticles in final-state.
To see how these cuts affect the signal and the background, the values
of selection efficiency are calculated and the invariant mass distributions
were drawn. Finally the statistical significance values of the expected
signal yield were calculated using different integrated luminosity
values within the range $L$ $\in$ $[1,100]$ $fb^{-1}$ with step
of $10$ $fb^{-1}$. The curves indicating the mass limits for exclusion,
observation and discovery of the excited electrons have been plotted
in the parameter space ($L$, $m^{\star}$). For example, mass limit
of observation of the excited electron at the PWFALC-SPPC2 collider
that has the highest center-of-mass energy is $16.8$ TeV for the
integrated luminosity $L=$$100$ $fb^{-1}$. Smaller mass values
are excluded.

As a result, the future SPPC-based lepton-hadron colliders will play
an important role in the investigation of the excited electrons.

\subsection*{Conflict of Interest}

The author declares that he has no conflict of interest.
\begin{acknowledgments}
I would like to thank Dr. A. Ozansoy for support of the model file.
This work has been supported by the Scientific and Technological Research
Council of Turkey (TÜB\.{I}TAK) under the Grant no. 114F337.\end{acknowledgments}

\end{document}